\documentclass{ws-mpla}

\begin{document}

\renewcommand{\draftnote}{}
\renewcommand{\trimmarks}{}

\markboth{E. A. Matute} {Neutrino mass generation with extra
right-handed fields}

\catchline{}{}{}{}{}

\title{Neutrino mass generation with extra right-handed fields\\
in a Dirac scenario via the type-I seesaw mechanism}

\author{\footnotesize Ernesto A. Matute}

\address{Departamento de F\'{\i}sica, Universidad de Santiago de
Chile,\\ Usach, Casilla 307 -- Correo 2, Santiago, Chile\\
ernesto.matute@usach.cl}

\maketitle

\pub{}{}

\begin{abstract}
An extension of the Standard Model (SM) is studied in which two
right-handed (RH) neutrinos per generation are incorporated, but
considering the hypothesis of the symmetry of lepton and quark
contents in order to deprive the number of RH neutrinos of
freedom, generate Dirac neutrinos and accommodate naturally tiny
values for their masses. The high scale type-I seesaw regime is
applied to the first, ordinary RH neutrino, whereas a low scale
pseudo-Dirac scenario is used for the second, adulterant RH
neutrino, implying that the first RH neutrino decouples at the
high scale, while the second RH neutrino survives down to the low
scale to pair off in a Dirac-like form with the corresponding
left-handed (LH) neutrino. The small mass and couplings of this
extra RH neutrino are explained by means of the statement of the
symmetry of fermionic content, only regarded as a guideline to the
natural choice of parameters since it is not a proper symmetry in
the Lagrangian.

\end{abstract}

\ccode{PACS Nos.: 14.60.St, 14.60.Pq, 11.30.Hv, 11.30.Fs}

\section{Introduction} \label{Introduction}

The tiny mass of neutrinos implied by neutrino flavor oscillation
experiments is a clear indication of new physics beyond the
Standard Model (SM).\cite{SM1}$^{\mbox{--}}$\cite{SM3} Neutrinos
are massless in the SM because only left-handed (LH) neutrinos,
Higgs doublets and renormalizable terms are included. The simplest
possibility to generate neutrino masses is then via the
incorporation of three right-handed (RH) neutrinos, maintaining
the gauge and Higgs sectors of the SM. Moreover, the addition of
these particles restores the chiral partners of neutrinos omitted
by the SM. Allowing general couplings, such additional fermions
permit to introduce both Dirac neutrino mass terms which conserve
lepton number, and Majorana neutrino mass terms which violate
lepton number conservation but are not forbidden by the gauge
symmetry of the SM. The Dirac neutrino masses are assumed of the
order of the charged lepton masses, while the Majorana masses are
arbitrary, being unrelated to the electroweak scale.

There are two special limits that have been trending topics in the
literature, relying on the magnitudes of Majorana masses relative
to Dirac masses:

(a) The low scale pseudo-Dirac
limit,\cite{pseudoDirac1,pseudoDirac2} in which the Majorana mass
terms are assumed to be much smaller than the Dirac ones. It leads
to Majorana eigenstates which are paired up into almost Dirac
neutrinos, with tiny mass differences between the components of
each pair. In this scenario there is a small violation of the
lepton number conservation and a restoration of the symmetry
between leptons and quarks in the sense of having a spectrum with
equal numbers of LH and RH leptons and quarks in each generation.
The smallness of Majorana masses is naturally attributed to the
breaking of lepton number symmetry. However, there is no
explanation of the smallness of Dirac masses relative to those of
charged leptons. Moreover, the expected rich low-energy
phenomenology of neutrino oscillations is essentially excluded by
experiments. Thus only the Dirac limit is allowed from the
viewpoint of phenomenology, although in this case the lepton
number symmetry that forbids Majorana masses has an ad-hoc
character.

(b) The high scale type-I seesaw
limit,\cite{highseesaw1}$^{\mbox{--}}$\cite{highseesaw4} in which
Majorana masses are supposed to be much greater than Dirac masses.
The RH neutrinos are approximately Majorana mass eigenstates and
become decoupled from the light, mainly LH, Majorana states. This
scenario also repairs the asymmetry of the fermionic content of
the SM, but provides a natural explanation for the large
difference between neutrino and charged lepton masses. The lepton
number conservation is restored when the tiny neutrino masses
approach to zero. The possibility of a low scale seesaw regime has
also been explored,\cite{lowseesaw1,lowseesaw2} where data on
neutrino oscillations, charged lepton flavor violating processes
and electroweak precision measurements are used to constrain their
couplings.

Here the seesaw scenario for neutrino masses has been implemented
by introducing extra RH neutrinos, in addition to the three RH
states mentioned above. Models based on this so-called extended
seesaw scenario have been proposed to allow for light neutrinos
without inserting small mass scales, although using additional
symmetries to forbid Dirac and/or Majorana mass terms for the
extra RH
neutrinos.\cite{extseesaw1}$^{\mbox{--}}$\cite{extseesaw5}

The seesaw mechanism predicts that massive neutrinos are Majorana
fermions. However, this has been disfavored by recent experimental
investigations on neutrinoless double-beta
decay,\cite{2beta1}$^{\mbox{--}}$\cite{2beta3} the only feasible
physical process with the possibility of determining at present
the Majorana character of neutrinos. Although the issue on the
nature of light neutrinos is still not resolved, the no
observation of signals in the search for neutrino double-beta
decay would strengthen their Dirac character, i.e. lighter
neutrinos can be Dirac particles like charged leptons and quarks.

Our aim in this paper is to study an extended mass model with
general couplings in which two RH neutrinos per generation of
leptons and quarks are incorporated, giving place for a general
mass matrix structure. Our main motivation is to generate Dirac
neutrinos and accommodate naturally tiny values for their masses
in a minimal extension of the SM, involving the popular high scale
type-I seesaw scenario on the one hand and the low scale
pseudo-Dirac scenario on the other hand. The hypothesis of the
symmetry of lepton and quark contents is used to deprive the
number of RH neutrinos of freedom and the seesaw mechanism is
applied to allow neutrinos having small masses and appearing to
have a Dirac nature, with a parameter region not excluded by
experiments. At this point we stress that the symmetry of
fermionic content is actually a lepton--quark correspondence but
not a symmetry in the Lagrangian of the model, which means that in
the electroweak sector of the SM extended with RH neutrinos one
cannot define a set of transformations between leptons and quarks
that keeps the Lagrangian invariant. Yet, this correspondence of
contents may serve as a guideline to the natural choice of
parameters leading to Dirac-like neutrinos with small masses,
expecting that further studies can attach it a proper symmetry but
in a different context. We do not address here aspects related
with leptonic mixing angles.

The work is organized as follows. In Sec.~2 we describe the
neutrino mass model in the simple case of one generation,
extending the results to three families in Sec.~3. In Sec.~4 we
consider the effective model at low energies. Phenomenological
remarks are given in Sec.~5. In Sec.~6 we summarize our
conclusions.

\section{Neutrino Masses with One Left-Handed and Two Right-Handed
Neutrinos} \label{one-family}

We first consider the extended scenario in the simplest case of
just one generation of neutrinos, so paving the way for the
three-generation extended model to be treated in the next section.
Two RH neutrinos are added to the SM in the approximation of one
generation, preserving its gauge and Higgs structure, i.e. only
one doublet of Higgs fields. The first RH neutrino is the ordinary
one denoted as $\nu_{R}$, which may carry a $B-L$ charge and form
a doublet with its RH charged lepton partner $e_{R}$, as in models
of left--right
symmetry.\cite{LRmodel1}$^{\mbox{--}}$\cite{LRmodel3} The other,
denoted as $\nu^{\prime}_{R}$, is a secondary singlet with small
couplings in comparison to the ones of $\nu_R$. Invoking the
't~Hooft's criterion,\cite{tHooft} this smallness appears natural
since a symmetry of lepton and quark contents is reestablished if
these couplings are set to be zero. In particular, we start taking
a light Majorana mass $m^{\prime}_R$ for $\nu^{\prime}_{R}$, and
assuming a heavy $m_R$ for $\nu_{R}$ as in the canonical high
scale type-I seesaw scenario. The question, however, is if the
symmetry of lepton and quark contents is good enough to ensure the
naturalness of the values chosen for the parameters of the model.
As noted in Sec.~\ref{Introduction}, our assumption is that at
least it serves, invoking the 't~Hooft's argument for small
numbers in the Lagrangian, as a guideline to the selection of
parameters, although the lepton--quark correspondence should have
attached a proper symmetry in a Lagrangian somehow connected with
the SM extended with RH neutrinos, which goes beyond the scope of
this work.

Following the notation of Ref.~\refcite{Langacker}, the Yukawa
Lagrangian containing the RH neutrinos $\nu_{R}$ and
$\nu^{\prime}_{R}$ and expanded with their respective Majorana
mass terms, becomes
\begin{eqnarray}
{\cal L} &=& - y_\nu \bar{L} \tilde{\phi} \nu_R - y^{\prime}_\nu
\bar{L} \tilde{\phi} \nu^{\prime}_R - \frac{1}{2} m_R
\bar{\nu}^c_L \nu_R - \frac{1}{2} m^\prime_R \bar{\nu}^{\prime
c}_L \nu^\prime_R \nonumber \\ && - \frac{1}{2} \mu^\prime
\bar{\nu}^{\prime c}_L \nu_R - \frac{1}{2} \mu^\prime
\bar{\nu}^c_L \nu^\prime_R + h.c. , \label{Yukawa}
\end{eqnarray}
where $L$ and $\phi$ are the lepton and Higgs doublets, $y_\nu$
and $y^\prime_\nu$ are the Yukawa couplings, the mixing term
$\mu^\prime$ of $\nu^\prime_R$ to $\nu_R$ is allowed, and
$\nu^c_R=C\bar{\nu}^T_L$. The classical mass terms after
spontaneous electroweak breaking can be written as
\begin{equation}
- {\cal L}_\nu = \frac{1}{2} \left( \begin{array}{ccc} \bar{\nu}_L
& \bar{\nu}^c_L & \bar{\nu}^{\prime c}_L \end{array} \right)
\left(
\begin{array}{ccc}
0 & m_D & m^\prime_D \\ m_D & m_R & \mu^\prime \\
m^\prime_D & \mu^\prime & m^\prime_R
\end{array} \right)
\left( \begin{array}{c} \nu^c_R \\ \nu_R \\ \nu^\prime_R
\end{array} \right) + h.c. ,
\label{massL}
\end{equation}
where $m_D=y_\nu \langle \phi^0 \rangle$ and
$m^\prime_D=y^\prime_\nu \langle \phi^0 \rangle$ refer to the
Dirac mass terms.

The masses and couplings of RH neutrinos should be fixed. Since
the origin of the phenomenological SM itself is even unknown, this
specification could not be expressed in a well defined form. Here
we follow the arguments of Shaposhnikov in favor of the hypothesis
of a lepton--quark symmetry regarding the particle content (see
Ref.~\refcite{Shapo} and references therein). At the level of the
SM there is an asymmetry between leptons and quarks: every LH
charged lepton and quark has its RH charged lepton or quark
partner, while the RH partner of the neutrino is absent. The
introduction of one RH neutrino, say $\nu_R$, simply reestablishes
the symmetry between leptons and quarks.\cite{Branco} Within the
context of Eqs.~(\ref{Yukawa}) and (\ref{massL}), it is given by
$m^\prime_D=0$ (or $y^\prime_\nu=0$), $\mu^\prime=0$ and
$m^\prime_R=0$, so that only $m_D$ and $m_R$ are different from
zero. Here our proposal takes the logic of the type-I seesaw
mechanism: It is natural to have $m_D$ of the same order of the
magnitude as charged leptons or quarks, and then $m_R$
sufficiently large to suppress $m_D$ according to $m^2_D/m_R$.

The inclusion of a second RH neutrino, $\nu^\prime_R$, breaks such
a lepton--quark correspondence. This is regarded as a reason for
having small couplings $m^\prime_D$, $\mu^\prime$, $m^\prime_R$
for $\nu^\prime_R$ in comparison to $m_D$, $m_R$ of $\nu_R$, as
the 't~Hooft's naturalness criterion applied to this symmetry of
lepton and quark contents in the Lagrangian gives a ready
explanation. It can be said that this extra RH neutrino sets an
alternative lepton--quark symmetry, but very weakened. Thus, the
lepton--quark symmetry distinguishes $\nu_R$ from $\nu^\prime_R$
by requiring a large difference between $m_D$, $m_R$ and
$m^\prime_D$, $m^\prime_R$, respectively, which parameterize the
two forms of the symmetry of fermionic content. Here it is worth
emphasizing that this symmetry of particle content cannot be
conceived as a symmetry of the electroweak Lagrangian under
transformations on the lepton and quark fields, because these have
different hypercharges and the Majorana mass terms for RH
neutrinos do not have counterparts in the quark sector. Since
everything may not be understood yet, our assertion is that the
't~Hooft's naturalness criterion as a guide of model construction
can be used in this case. In the following we show that a soft
breaking of the correspondence between leptons and quarks, stated
by the hypothesis of the symmetry of fermionic content, through
extra RH neutrinos can lead to light neutrinos of Dirac type,
where the questioned Majorana mass terms are suppressed or
decoupled from the low-energy effective model. It can be seen that
the high scale type-I seesaw is mimicked with $m^\prime_D =
\mu^\prime = m^\prime_R = 0$, while the low scale pseudo-Dirac
neutrino pair is done with $m_D = \mu^\prime = m_R = 0$.

With all of the above in mind, we assume $m^\prime_R, \mu^\prime,
m^\prime_D, m_D \ll m_R$, extending the high scale seesaw
scenario. In this limit we would anticipate suppressions of the
Dirac mass $m_D$ and the coupling $\mu^\prime$ according to
$m^2_D/m_R$ and $\mu^{\prime 2}/m_R$, respectively. Also, now
following our motivations stated in Sec.~1, in order to have a low
scale pseudo-Dirac regime we assume the inequalities $m^\prime_R$,
$\mu^{\prime 2}/m_R$, $m^2_D/m_R$, $m_D \mu^\prime/m_R \ll
m^\prime_D$. As a matter of fact, in the case of these mass
hierarchies we obtain, by applying directly the Cardano's formula
for the roots of a cubic equation, the mass eigenvalues
\begin{eqnarray}
\begin{array}{l}
\displaystyle m_1 \simeq - m^\prime_D + \frac{1}{2} m^\prime_R -
\frac{1}{2} \frac{(m_D-\mu^\prime)^2}{m_R} \simeq - m^\prime_D ,
\\ [10pt] m_2 \simeq m_R , \\ [10pt] \displaystyle m_3 \simeq
m^\prime_D + \frac{1}{2} m^\prime_R - \frac{1}{2}
\frac{(m_D+\mu^\prime)^2}{m_R} \simeq m^\prime_D ,
\end{array}
\label{masses}
\end{eqnarray}
where only the leading terms in $m^\prime_D$, $m^\prime_R$,
$\mu^\prime$, $m_D$, and $m_R$ are shown. As expected,
$m^\prime_D$ and $m^\prime_R$ are not suppressed by $m_R$. We find
that the mass matrix is diagonalized by the approximately unitary
matrix
\begin{eqnarray}
{\cal U}^\dagger &\simeq& \left( \begin{array}{ccc} \displaystyle
\left( \frac{1}{\sqrt{2}}+w \right) \; & \displaystyle \left[
-\frac{m_D}{m_R} \left( \frac{1}{\sqrt{2}}+w \right)
+\frac{\mu^\prime}{m_R} \left( \frac{1}{\sqrt{2}}-w \right)
\right] \; & \displaystyle \left( -\frac{1}{\sqrt{2}}+w \right) \\
[10pt] \displaystyle \frac{m_D}{m_R} & 1 & \displaystyle
\frac{\mu^\prime}{m_R} \\ [10pt] \displaystyle \left(
\frac{1}{\sqrt{2}}-w \right) & \displaystyle \left[
-\frac{m_D}{m_R} \left( \frac{1}{\sqrt{2}}-w \right)-
\frac{\mu^\prime}{m_R} \left( \frac{1}{\sqrt{2}}+w \right) \right]
& \displaystyle \left( \frac{1}{\sqrt{2}}+w \right)
\end{array} \right) \nonumber \\ && \nonumber \\
&\simeq& \left( \begin{array}{clc} \displaystyle
\frac{1}{\sqrt{2}} \;\;\; & 0 \;\; & \displaystyle
-\frac{1}{\sqrt{2}}
\\ [10pt] 0 & 1 & 0 \\ [10pt] \displaystyle \frac{1}{\sqrt{2}}
& 0 & \displaystyle \frac{1}{\sqrt{2}}
\end{array} \right) ,
\end{eqnarray}
where
\begin{equation}
w = \frac{1}{4\sqrt{2}} \frac{m^\prime_R}{m^\prime_D} +
\frac{1}{4\sqrt{2}} \frac{m^2_D - \mu^{\prime 2}}{m_R m^\prime_D}
, \label{w}
\end{equation}
so that
\begin{equation}
{\cal U}^\dagger {\cal M} {\cal U}^{*} = \left(
\begin{array}{ccc} m_1 & 0 & 0 \\ 0 & m_2 & 0 \\ 0 & 0 & m_3
\end{array} \right) ,
\end{equation}
with ${\cal M}$ being the symmetric mass matrix of
Eq.~(\ref{massL}). The mass eigenvalues can be made positive by
suitable phase choice in the chiral fields.

Thus the state $\nu_R$, the natural partner of $\nu_L$,
approximately becomes mass eigenstate and is decoupled at low
energy. The LH state then combines with almost maximal mixing with
the secondary RH state, its unconventional partner whose mass
couplings are relatively small. Specifically, the mass eigenstates
correspond to the three Majorana combinations given by $\nu_{iM} =
\nu_{iL}+\nu^c_{iR}$, with $i = 1, 2, 3$, related to the weak
states through the transformations
\begin{eqnarray}
\begin{array}{l}
\left( \begin{array}{c} \nu_{1L} \\ \nu_{2L} \\ \nu_{3L}
\end{array} \right) = {\cal U}^\dagger \left( \begin{array}{c}
\nu_{L} \\ \nu^c_L \\ \nu^{\prime c}_L \end{array} \right) \simeq
\left( \begin{array}{c}
\frac{1}{\sqrt{2}}(\nu_{L}-\nu^{\prime c}_L) \\ \nu^c_L \\
\frac{1}{\sqrt{2}}(\nu_L+\nu^{\prime c}_L) \end{array} \right) ,
\\ [30pt]
\left( \begin{array}{c} \nu^c_{1R} \\ \nu^c_{2R} \\ \nu^c_{3R}
\end{array} \right) = {\cal U}^T \left( \begin{array}{c}
\nu^c_{R} \\ \nu_R \\ \nu^\prime_R \end{array} \right) \simeq
\left( \begin{array}{c}
\frac{1}{\sqrt{2}}(-\nu^\prime_R+\nu^c_R) \\ \nu_R \\
\frac{1}{\sqrt{2}}(\nu^\prime_R+\nu^c_R) \end{array} \right) .
\end{array}
\end{eqnarray}

Clearly, there is a suppression of the mixing of the neutrino
$\nu_L$ with its ordinary partner $\nu_R$, and a suppression of
the usual Dirac mass $m_D$ relative to $m^\prime_D$. This
situation leads to an almost degenerate pair of eigenstates with a
small mass difference given by $\Delta m \simeq
|m^\prime_R-(m^2_D+\mu^{\prime2})/m_R| \ll m^\prime_D$.

Now, since $m^\prime_R$ is not suppressed and not needed as
another small mass scale, we set $m^\prime_R=0$ and the
pseudo-Dirac regime may proceed via the suppressed terms
containing $m_D$ and $\mu^\prime$. This is equivalent to
effectively having a lepton number conservation at low energies,
assuming a high seesaw scale.

Within the standard pseudo-Dirac framework, with only one RH
neutrino, a small value for the Dirac neutrino mass $m_D$ is
considered unnatural. In our extended pseudo-Dirac scenario,
however, a small Dirac neutrino mass $m^\prime_D$ becomes natural
in the sense of 't Hooft\cite{tHooft} because a symmetry of lepton
and quark contents is restored if the mixing couplings of the
adulterant state $\nu^\prime_R$ vanish (see Eq.~(\ref{Yukawa})).
Again, as stressed above, this lepton--quark correspondence only
serving as a guideline to the choice of parameters.

\section{Extension to Three Generations of Neutrinos}

We now generalize the results of Sec.~2 to the more realistic
scenario of three generations of LH neutrinos. The particle
content of the SM is augmented by two RH neutrinos per generation.
The three LH neutrinos $\nu_L$, the three ordinary RH neutrinos
$\nu_R$, and the three adulterant RH neutrinos $\nu^\prime_R$ have
mass terms that can be written in a form similar to
Eq.~(\ref{massL}), with the mass matrix replaced by
\begin{equation}
{\cal M} = \left(
\begin{array}{ccc}
0 & M_D & M^\prime_D \\ M^T_D & M_R & M^{\prime T} \\
M^{\prime T}_D & M^\prime & M^\prime_R
\end{array} \right) ,
\end{equation}
where $M_R$, $M^\prime_R$, $M_D$, $M^\prime_D$, and $M^\prime$ are
3$\times$3 complex matrices. It can be diagonalized by the unitary
transformation
\begin{equation}
{\cal U}^\dag {\cal M} {\cal U}^{*} = \left(
\begin{array}{ccc} D_L & 0 & 0 \\ 0 & D_R & 0 \\ 0 & 0 &
D^\prime_R \end{array} \right) ,
\end{equation}
where $D_L$, $D_R$ and $D^\prime_R$ are diagonal, real and
non-negative 3$\times$3 matrices. We consider
\begin{equation}
{\cal U}^\dagger = \left( \begin{array}{ccc} V^\dagger_L & 0 & 0
\\ 0 & V^\dagger_R & 0 \\ 0 & 0 & V^{\prime \dagger}_R \end{array}
\right) \left(
\begin{array}{ccc}
\frac{1}{\sqrt{2}}I+W^\dagger_{LL} \;\;\;\; & V^\dagger_{RL}
\;\;\;\; & -\frac{1}{\sqrt{2}}I+W^{\prime \dagger}_{RL} \\
V^\dagger_{LR} & I & V^{\prime \dagger}_{RL} \\
\frac{1}{\sqrt{2}}I+W^{\prime \dagger}_{LR} & V^{\prime
\dagger}_{LR} & \frac{1}{\sqrt{2}}I+W^{\prime \dagger}_{RR}
\end{array} \right) ,
\end{equation}
where $V_L$, $V_R$ and $V^\prime_R$ are unitary 3$\times$3 complex
matrices. Assuming that $M_R$ and $M^\prime_D$ are nonsingular and
symmetric matrices, and that $M^\prime_R$, $M^\prime$,
$M^\prime_D$, $M_D \ll M_R$ as well as $M^\prime_R$, $M_D M^{-1}_R
M^T_D$, $M^\prime M^{-1}_R M^{\prime T}$, $M^\prime M^{-1}_R M^T_D
\ll M^\prime_D$, we use the constraints from unitarity and the
matrix ${\cal MU}^{*}$ as in the ordinary seesaw mechanism to
finally obtain:
\begin{eqnarray}
\begin{array}{l}
W^\dagger_{LL} \simeq \frac{1}{4\sqrt{2}} M^\prime_R M^{\prime
-1}_D + \frac{1}{4\sqrt{2}} (M_D-M^\prime) M^{-1}_R
(M^T_D+M^{\prime T}) M^{\prime -1}_D , \\ [10pt] W^{\prime
\dagger}_{RR} \simeq \frac{1}{4\sqrt{2}} M^\prime_R M^{\prime
-1}_D + \frac{1}{4\sqrt{2}} (M_D+M^\prime) M^{-1}_R
(M^T_D-M^{\prime T}) M^{\prime -1}_D , \\ [10pt] W^{\prime
\dagger}_{RL} \simeq W^\dagger_{LL} , \\ [10pt] W^{\prime
\dagger}_{LR} \simeq - W^{\prime \dagger}_{RR} , \\ [10pt]
V^\dagger_{RL} \simeq - (\frac{1}{\sqrt{2}}I+W^\dagger_{LL}) M_D
M^{-1}_R + (\frac{1}{\sqrt{2}}I-W^\dagger_{LL}) M^\prime M^{-1}_R
, \\ [10pt]  V^{\prime \dagger}_{LR} \simeq -
(\frac{1}{\sqrt{2}}I-W^{\prime \dagger}_{RR}) M_D M^{-1}_R -
(\frac{1}{\sqrt{2}}I+W^{\prime \dagger}_{RR}) M^\prime M^{-1}_R ,
\\ [10pt] V^\dagger_{LR} \simeq M^{-1 \dagger}_R M^\dagger_D
, \\ [10pt] V^{\prime \dagger}_{RL} \simeq M^{-1 \dagger}_R
M^{\prime \dagger} .
\end{array}
\label{Ws}
\end{eqnarray}
Thus, we get
\begin{eqnarray}
D_L &\simeq& V^\dagger_L [-M^\prime_D+\frac{1}{2} M^\prime_R -
\frac{1}{2} (M_D-M^\prime)M^{-1}_R(M^T_D-M^{\prime T})] V^{*}_L
\nonumber \\ &\simeq& - V^\dagger_L M^\prime_D V^{*}_L , \nonumber \\
D_R &\simeq& V^\dagger_R M_R V^{*}_R , \label{Ds} \\
D^\prime_R &\simeq& V^{\prime \dagger}_R [M^\prime_D+\frac{1}{2}
M^\prime_R - \frac{1}{2} (M_D+M^\prime)M^{-1}_R(M^T_D+M^{\prime
T})] V^{\prime *}_R \nonumber \\ &\simeq& V^{\prime \dagger}_R
M^\prime_D V^{\prime *}_R . \nonumber
\end{eqnarray}

In the pseudo-Dirac limit with $M^\prime_R=0$ and $M_D$,
$M^\prime$ suppressed, there are three light almost degenerate
pairs of mass eigenstates with small mass differences, with almost
maximal mixing of LH neutrinos $\nu_L$ and adulterant RH neutrinos
$\nu^\prime_R$, and three heavy, mostly ordinary RH neutrinos
$\nu_R$ with mass matrix $M_R$. The masses of light neutrinos are
of the order of $M^\prime_D$ instead of $M_D$, which are
suppressed by the seesaw mechanism. The matrices $V_{LR}$,
$V_{RL}$, $V^\prime_{LR}$ and $V^\prime_{RL}$ are suppressed by
$M_R$, whereas $W_{LL}$, $W^\prime_{RR}$, $W^\prime_{LR}$ and
$W^\prime_{RL}$ are suppressed by $M_R$ and/or $M^\prime_D$. We
note that the results in Eqs.~(\ref{Ws}) and (\ref{Ds}) reproduce
those obtained in Sec.~2 in the case of one generation, which were
calculated following a completely different method.

\section{Low-Energy Effective Model of Neutrino Masses}

The RH neutrinos with huge masses can be integrated out using the
equation of motion
\begin{equation}
\frac{d {\cal L}_\nu}{d \nu_R} = 0 . \label{motionEq}
\end{equation}
In the approximation of one generation, it leads to
\begin{equation}
\bar{\nu}^c_L = - \frac{m_D}{m_R} \bar{\nu}_L -
\frac{\mu^\prime}{m_R} \bar{\nu}^{\prime c}_L , \qquad \nu_R = -
\frac{m_D}{m_R} \nu^c_R - \frac{\mu^\prime}{m_R} \nu^\prime_R .
\end{equation}
The effective Lagrangian we then have is
\begin{equation}
- {\cal L}_\nu = \displaystyle \frac{1}{2} \left(
\begin{array}{cc} \bar{\nu}_L \; & \bar{\nu}^{\prime c}_L
\end{array} \right) \left(
\begin{array}{cc} \displaystyle
-\frac{m^2_D}{m_R} \; & \displaystyle m^\prime_D-\frac{\mu^\prime
m_D}{m_R} \\ & \\ \displaystyle m^\prime_D-\frac{\mu^\prime
m_D}{m_R} & \displaystyle -\frac{\mu^{\prime 2}}{m_R}
\end{array} \right)
\left( \begin{array}{c} \nu^c_R \\ \\ \nu^\prime_R
\end{array} \right) + h.c.,
\end{equation}
where $m^\prime_R=0$ is used, so that the pseudo-Dirac scenario
proceeds via suppressed mass terms, without the need of inserting
a second small mass scale. The mass matrix is diagonalized by the
approximately unitary matrix
\begin{equation}
{\cal U} \simeq \left( \begin{array}{cc} \frac{1}{\sqrt{2}}+w
\;\;\;\;\; & \frac{1}{\sqrt{2}}-w \\ & \\ -\frac{1}{\sqrt{2}}+w &
\frac{1}{\sqrt{2}}+w
\end{array} \right) ,
\end{equation}
such that
\begin{equation}
{\cal U}^\dagger {\cal M} {\cal U}^{*} = \left(
\begin{array}{cc} m_1 & 0 \\ 0 & m_3
\end{array} \right) ,
\end{equation}
where $w$, $m_1$ and $m_3$ are given in Eqs.~(\ref{w}) and
(\ref{masses}), with $m^\prime_R=0$.

On the other hand, assuming the mass hierarchy
\begin{equation}
\frac{m^2_D}{m_R}, \frac{\mu^\prime m_D}{m_R}, \frac{\mu^{\prime
2}}{m_R} \ll m^\prime_D \ll m_D \ll m_R ,
\end{equation}
we end up with the mass matrix
\begin{equation}
{\cal M} \simeq \left(
\begin{array}{cc} 0 & m^\prime_D \\ m^\prime_D & 0
\end{array} \right) .
\end{equation}
This result is consistent with a generation of standard leptons
$L=(\nu_L,e_L)$, $e_R$ extended with the extra RH neutrino
$\nu^\prime_R$ and a Lagrangian which includes the Yukawa terms
related to $\nu^\prime_R$,
\begin{equation}
{\cal L} = - y^{\prime}_\nu \bar{L} \tilde{\phi} \nu^{\prime}_R +
h.c. \label{Y-terms}
\end{equation}

In the limit in which the small values $m^2_D /m_R$, $\mu^\prime
m_D / m_R$, $\mu^{\prime 2} / m_R$ are equal to zero, a lepton
number conservation and a lepton--quark symmetry are set up at low
energies, below the mass scale of the ordinary RH neutrino
$\nu_R$. It is the lepton--quark symmetry in terms of
$\nu_R^\prime$ defined in Sec.~\ref{one-family}, with all
couplings of $\nu_R$ removed ($m_D=m_R=\mu^\prime=0$). Now, a
neutrino Dirac mass $m^\prime_D$ much smaller than $m_D \sim m_e$
appears natural because $m^\prime_D=0$ (with
$\mu^\prime=m^\prime_R=0$) recovers an enhanced symmetry in the
original Lagrangian, namely, the symmetry of lepton and quark
contents involving the natural neutrino partner $\nu_R$. Thus
light Dirac neutrinos with small masses or Yukawa couplings may be
accommodated naturally, as written in Eq.~(\ref{Y-terms}),
although the arguments are based on the correspondence between
lepton and quark contents which is merely a guideline to the
choice of parameters and not a proper symmetry in the Lagrangian,
as emphasized above. It appears as an alternative to the usual
approach which extends the SM with the Yukawa terms ${\cal L} = -
y_\nu \bar{L} \tilde{\phi} \nu_R$ in order to have Dirac
neutrinos.

The above results can be generalized to three generations.
Equation (\ref{motionEq}) now leads to
\begin{equation}
\bar{\nu}^c_L = - \bar{\nu}_L M_D M^{-1}_R - \bar{\nu}^{\prime
c}_L M^\prime M^{-1}_R , \hspace{20pt} \nu_R = - M^{-1}_R M^T_D
\nu^c_R - M^{-1}_R M^{\prime T} \nu^\prime_R .
\end{equation}
The effective Lagrangian is written as
\begin{equation}
- {\cal L}_\nu = \frac{1}{2} \left( \begin{array}{cc} \bar{\nu}_L
& \bar{\nu}^{\prime c}_L \end{array} \right) \left(
\begin{array}{ll}
M_{LL} \;\; & M^\prime_{LR} \\ M^{\prime T}_{LR} & M^\prime_{RR}
\end{array} \right)
\left( \begin{array}{c} \nu^c_R \\ \nu^\prime_R
\end{array} \right) + h.c. , \label{effL}
\end{equation}
where
\begin{eqnarray}
\begin{array}{c}
M_{LL} \simeq - M_D M^{-1}_R M^T_D , \qquad M^\prime_{LR} \simeq
M^\prime_D - M_D M^{-1}_R M^{\prime T} , \\ [10pt] M^\prime_{RR}
\simeq - M^\prime M^{-1}_R M^{\prime T} .
\end{array}
\end{eqnarray}

The mass matrix of Eq.~(\ref{effL}) is diagonalized by the
approximately unitary matrix
\begin{equation}
{\cal U}^\dagger \simeq \left( \begin{array}{cc} V^\dagger_L & 0
\\ & \\ 0 & V^{\prime \dagger}_R \end{array} \right) \left(
\begin{array}{lc}
\frac{1}{\sqrt{2}}I+W^\dagger_{LL} \;\; &
-\frac{1}{\sqrt{2}}I+W^\dagger_{LL} \\ & \\
\frac{1}{\sqrt{2}}I-W^{\prime \dagger}_{RR} &
\frac{1}{\sqrt{2}}I+W^{\prime \dagger}_{RR}
\end{array} \right) ,
\end{equation}
such that
\begin{equation}
{\cal U}^\dag {\cal M} {\cal U}^{*} = \left(
\begin{array}{cc} D_L & 0 \\ 0 & D^\prime_R \end{array}
\right) ,
\end{equation}
where $W^\dagger_{LL}$ and $W^{\prime \dagger}_{RR}$, $D_L$ and
$D^\prime_R$ have the expressions given in Eqs.~(\ref{Ws}) and
(\ref{Ds}) with $M^\prime_R=0$.

\section{Phenomenological Remarks}

We have considered an scenario where each LH neutrino $\nu_L$ has
two RH partners: $\nu_R$ and $\nu^\prime_R$. A seesaw mechanism of
type-I has been applied in which the Majorana mass $m_R$ of
$\nu_R$ is assumed to be much larger than the Dirac mass $m_D$
coupling $\nu_L$ to $\nu_R$. The state $\nu_R$ is decoupled at low
energies leaving $\nu^\prime_R$ as the main partner of $\nu_L$.
The mass $m_D$ is assumed to be of order the charged lepton mass.
In the approximation of one generation, $m_D \sim m_e \sim$ 1 MeV,
while the mass term $m_R$ may be as large as the scale of Grand
Unification Theories, say $m_R \sim 10^{14}$ GeV, and in principle
even up to the Planck mass. This would leads to an effective LH
Majorana mass of order
\begin{equation}
m_{LL} = \frac{m^2_D}{m_R} \sim 10^{-11} \, \mbox{eV} .
\end{equation}

On the other hand, it is found that oscillations of solar
neutrinos set an upper bound for $m^\prime_{RR}$,\cite{mRR}
\begin{equation}
m^\prime_{RR} = m^\prime_R - \frac{\mu^{\prime 2}}{m_R} \lesssim
10^{-9} \, \mbox{eV} .
\end{equation}
Next, taking from the neutrino data\cite{PDG}
\begin{equation}
m^\prime_{LR} = m^\prime_D - \frac{\mu^\prime m_D}{m_R} \sim
10^{-1} \, \mbox{eV} ,
\end{equation}
we have the following benchmark values for the parameters in the
model,
\begin{eqnarray}
\begin{array}{c}
m_R \sim 10^{14} \, \mbox{GeV} , \qquad m_D \sim 1 \, \mbox{MeV} , \\
[10pt] \mu^\prime \lesssim 10 \, \mbox{MeV} , \qquad m^\prime_R
\lesssim 10^{-9} \, \mbox{eV} , \qquad m^\prime_D \sim 10^{-1} \,
\mbox{eV} ,
\end{array}
\end{eqnarray}
with the expected hierarchy of masses
\begin{equation}
m_{LL} , m^\prime_{RR} \ll m^\prime_{LR} \ll m_D \ll m_R ,
\end{equation}
so realizing the approximations used in the model, where in the
end light neutrinos appear to have a Dirac character.

The phenomenological implications at low energies are essentially
those of the usual Dirac approach, while at high energies the
model maintains the expectations of the high scale type-I seesaw
mechanism.\cite{highseesaw1}$^{\mbox{--}}$\cite{highseesaw4} The
parameter region we have considered is consistent with
experimental bounds which exclude the pseudo-Dirac limit, but not
a Dirac nature for light neutrinos. And their masses or Yukawa
couplings may have exceptionally small values because of the
adulterant character of RH partners. Besides, there is consistency
between this Dirac picture and the vanishing of the Majorana mass
$m^\prime_R$ assumed above. Also, the Dirac nature of lighter
neutrinos, as effectively implied in this work, refuses to allow
the neutrinoless double-beta decay, in accordance with recent
precision experiments.\cite{2beta1}$^{\mbox{--}}$\cite{2beta3}

\section{Conclusions}

We have constructed an extension of the SM by incorporating two RH
neutrinos per generation of leptons and quarks, but considering
the hypothesis of the symmetry of fermionic content in order to
deprive the number of RH neutrinos of freedom, generate Dirac
neutrinos and accommodate naturally their tiny masses. One of
these is the ordinary RH neutrino which restores the
correspondence between leptons and quarks at high energies with
weak couplings having order of magnitudes as those of its weak
charged partner and a Majorana mass term whose coupling is assumed
to be large, as in the canonical high scale type-I seesaw
scenario. The other, adulterant, RH neutrino, which breaks the
lepton--quark symmetry established with the first one, is regarded
to have relatively small mass and couplings, as the 't~Hooft's
naturalness criterion applied to this symmetry of lepton and quark
contents provides a ready explanation. The first RH neutrino is
decoupled at the high scale, but the second RH neutrino survives
down to the low scale to pair off in a Dirac-like fashion with the
corresponding LH neutrino, imposing its own form of the symmetry
of fermionic content.

We have emphasized, however, that the correspondence of lepton and
quark contents is not an actual symmetry in the Lagrangian because
one cannot write a symmetry transformation between leptons and
quarks to keep the Lagrangian invariant. As it is well-known, the
't~Hooft's argument for small parameters in a Lagrangian relies on
the symmetry, which guarantees the quantum corrections of such
numbers to be proportional to the parameters themselves. Its
application to the lepton--quark correspondence therefore demands
the attachment of a proper symmetry in a Lagrangian somehow
associated with the SM extended with RH neutrinos, which surpasses
the aims of this work. Yet, we have considered that it serves as a
guideline to the natural choice of parameters of small values.

Thus, a low scale Dirac scenario with lepton--quark symmetry of
content and small neutrino masses appears to be natural with extra
RH neutrinos via the high scale type-I seesaw mechanism. The
parameter region considered in this approach makes irrelevant to
low energy processes the perturbation of the seesaw mechanism on a
description given in terms of light Dirac neutrinos, foreseeing
that experiments will not have sensitiveness to the Majorana
character of neutrinos predicted by the seesaw mechanism, as in
the case of the neutrinoless double-beta decay.

The usual Dirac scenario deals with the same chiral neutrino
included in the alternative seesaw mechanism, which generates
problems to explain naturally the smallness of Dirac neutrino mass
terms relative to those of charged leptons. Our key result making
the difference with this approach is that the symmetry of
fermionic content at low energies is achieved only with the
additional RH neutrinos and not with the ordinary ones which are
decoupled, may carry $B-L$ charge and form RH doublets with RH
leptons as in models of left--right symmetry. In other words,
light pseudo-Dirac neutrinos obtained by replacing the regular
factor $m_D$ by the new and independent, naturally small parameter
$m^\prime_D$, giving an understanding why observed neutrinos are
ultralight and at the same time Dirac-like. While there are no
hard predictions for the light neutrino masses and mixings, let
alone the mass hierarchy, this new framework opens up a new line
for future exploration.

\section*{Acknowledgments}

This work was supported by the Departamento de Investigaciones
Cient\'{\i}ficas y Tecnol\'ogicas, Universidad de Santiago de
Chile, Usach, Grant No. 041431MC.

\end{document}